\documentclass[prb,twocolumn,showpacs,preprintnumbers,amsmath,amssymb,floatfix]{revtex4}
\usepackage{graphicx}
\usepackage{float}
\usepackage{color}

\newcommand \be{\begin{eqnarray}}
\newcommand \ee{\end{eqnarray}}

\newcommand \ba{\begin{align}}
\newcommand \eea{\end{align}}

\begin{document}
           \csname @twocolumnfalse\endcsname
\title{Exact ground-state properties of one-dimensional electron gas at high density}
\author{Vinod Ashokan$^1$, Renu Bala$^{2}$, Klaus Morawetz$^{3,4}$\footnote{Corresponding author Email: morawetz@fh-muenster.de} and  Kare N. Pathak$^5$  }

\affiliation{$^1$Department of Physics, Dr. B.R. Ambedkar National
Institute of Technology, Jalandhar (Punjab) - 144 011, India}
\affiliation{$^2$Department of Physics, MCM DAV College for Women,
Chandigarh - 160036, India} \affiliation{$^3$M\"unster University
of Applied Sciences, Stegerwaldstrasse 39, 48565 Steinfurt,
Germany} \affiliation{$^4$International Institute of Physics-
UFRN, Campus Universit\'ario Lagoa nova, 59078-970 Natal, Brazil }
\affiliation{$^5$Centre for Advanced Study in Physics, Panjab
University, Chandigarh - 160014, India}
\begin{abstract}
The dynamical response theory is used to obtain an analytical
expression for the exchange energy of a quantum wire for arbitrary
polarization and width. It  reproduces the known form of exchange
energy for 1D electron gas in the limit of infinitely thin
cylindrical and harmonic wires. The structure factor for these
wires are also obtained analytically in the high-density or small
$r_s$ limit. This structure factor enables us to get the {\it
exact} correlation energy for both the wires and demonstrates that
there are at least two methods to get the ideal Coulomb limit in
one dimension. The structure factor and the correlation energy are
found to be independent of the way the one-dimensional Coulomb
potential is regularized. The analytical expression for the pair
correlation function is also presented for small distances and
provides a justification for the small $r_s$ expansion as long as
$r_s< \frac{3}{2} \left(\frac{\pi^2}{\pi^2+3} \right)=1.15$.
\end{abstract}
\maketitle

\section{Introduction}
The correlation energy for many-body electron system has drawn much attention \cite{Giuliani05,Giamarchi04} due to its utility and theoretically challenge. In particular, the one-spatial dimensional system is of current interest. Such one dimensional systems can be envisaged experimentally in carbon nanotubes \cite{Saito98,Bockrath99,Ishii03,Shiraishi03}, edge states in quantum hall liquid \cite{Milliken96,Mandal01,Chang03}, semiconducting nanowires \cite{Schafer08,Huang01}, cold atomic gases \cite{Monien98,Recati03,Moritz05} and conducting molecules \cite{Nitzan03}.

The correlation energy of a uniform electron gas is an important ingredient of most local and non-local density functional calculations. The Fermi liquid paradigm works well for interacting electrons in 2D, 3D systems, but it eventually breaks down in 1D due to the Peierls instability. However, the random phase approximation (RPA) is the correct theory in the large electron density regime \cite{KM18} i.e  $n=1/(2 r_s a_B)$, with $r_s$ being the coupling parameter and  $a_B$ the effective Bohr radius. The prospect to observe non-Fermi-liquid features has given a large impetus to both theoretical and experimental research. The physical properties of 1D interacting systems (Fermions, Bosons, spins) is described  theoretically by the Tomonaga-Luttinger liquid model\cite{Tomonaga50,Luttinger63,Haldane81}.

Recently we have reported the ground state properties of the 1D electron fluid at high density for an infinitely thin wire and a harmonic wire of finite thickness using a variational quantum Monte Carlo (QMC) method \citep{Vinod18c}. The simulation data of correlation energy (in unit of Hartree) for the infinitely thin wire is well represented by  $\epsilon_c(r_s)=-0.027431(3)+ 0.00791(1) r_s - 0.00196(1) r_s^2$. Furthermore the conventional perturbation theory \cite{Loos13,Loos16} gives the  expression for the correlation energy as  $\epsilon_c(r_s) = -\pi^2/360 + 0.00845 r_s + \dots$. The first and second term in the expression is due to the second order and third order perturbation theory respectively.

The purpose of the present paper is to study the ground-state properties of the interacting electron gas at high densities using a dynamical approach and to see what one can learn from this method. We will derive an analytical expression for the exchange energy of both cylindrical and harmonic wires of finite thickness $b$, both of which reduce to the same exchange energy in the limit of infinitely small thickness. Further the static structure factor is also obtained for both wires which also turns out to be the same in the limit $b\rightarrow 0$. Our analytical expressions for the exchange energy, the structure factor and the correlation energy for cylindrical and harmonic wires are respectively the same for $b\rightarrow 0$. This explicitly demonstrates that they are independent on the way the one-dimensional Coulomb potential is regularized at $x=0$. The high-density structure factor provides the exact analytical expression for the pair correlation function $g(r)$ for small $r$. This implies that the variational Monte Carlo wave function has nodes at the coalescence point $r=0$ which is in agreement with quantum Monte Carlo simulations. Further the structure factor enables us to get exactly the same correlation energy as obtained by the static perturbation theory \cite{Loos13} for $r_s\rightarrow 0$.

The paper is organized as follows. In section \ref{Theory}, first we describe the  regularization of the Coulomb potential by confining the electron through a harmonic potential and through any arbitrary confinement perpendicular to the axis of the wire (i.e cylindrical). In this section we also evaluate the density response function with  RPA including exchange. We derive the expression for exchange energy, structure factor, pair correlation function and correlation energy in section \ref{AnalyticalExpressions}. In section \ref{Conclusion} we summarize and conclude on the results.

\section{Theoretical formulation}
\label{Theory}

\subsection{Model potential}

The Fourier transform for the Coulomb potential $v(x) \varpropto1/x$ is  constant $-i \pi {\rm sgn} (k)$ and not a Coulumb one anywhere
in Fourier space. To avoid this divergence at small inter-electronic distances $x$, we model the interactions by a soften Coulomb potential of a cylindrical wire $V(x)=e^2/4 \pi \epsilon_0 \sqrt{x^2+\bar b^2}$ with the transverse width parameter $\bar b$ of the wire. Its Fourier transform reads
\be
V(q)&=&{e^2\over 4 \pi \epsilon_0} v(q)
\nonumber\\
v(q)&=&2 K_0(\bar b q)=
-2 \left[\ln \left(\frac{\bar b q}{2}\right)
+\gamma \right]
\nonumber\\
&&
-\frac{\bar b^2q^2}{2}
   \left[\ln \left(\frac{\bar b q}{2}\right)+\gamma
   -1\right]+O\left(\bar b^3\right)
\label{v}
\ee
where $K_{0}$ is the modified Bessel function of $2^{nd}$ kind.
The soften Coulomb potential for a harmonically trapped electron wire is given together with its Fourier transform
\be
V(r)&=&{e^2\over 4 \pi \epsilon_0}{\sqrt{\pi}\over b}{\rm e}^{x^2\over 4 b^2} {\rm erfc}\left ({|x|\over 4 b}\right )
\nonumber\\
v(q)&=&\frac{e^2}{\epsilon_{0}}~E_{1}(b^2 q^2)~ e^{b^2 q^2}
\ee
with  $E_{1}$ being the exponential integral.
The true long-range character of the Coulomb potential has been studied by Schulz \cite{Schulz93} and the mapping of long-range Coulomb interaction onto an exactly solvable model with short-range behavior has been studied by Fogler \cite{Fogler05a,Fogler05b}.
The calculation of the ground-state energy for thin wires in the high-density limit for realistic long-range Coulomb interactions is still an open problem for the 1D homogeneous electron gas.
To compare both inter-electronic interactions, the harmonic wire approaches
\be
 v(q)=\left\{%
\begin{array}{ll}
\label{HarPoten}
   -\gamma-2\ln (\bar b q) ~~\text{for} & \bar bq \rightarrow 0 \\
    (\bar b q)^{-2} ~~~~~~~~~~ \text{for} & \bar bq \rightarrow \infty ,\\
\end{array}%
\right.\ee
where $\gamma$ is the Euler constant and the cylindrical potential
\be
 v(q)=\left\{%
\begin{array}{lcl}
\label{ClyPoten}
   -2 \gamma-2\ln 2 -2\ln (\bar b q) &\text{for} & \bar bq \rightarrow 0 \\
   e^{-\bar bq}\sqrt{\frac{2\pi}{\bar bq}}&\text{for} & \bar bq \rightarrow \infty. \\
\end{array}%
\right.\ee
Both potentials behave similarly at the small $q$ limit, but at large $q$ they differ. It is noted that the harmonic potential is a Coulomb potential for both small and large $q$, however the large $q$ behavior of cylindrical potential is not a Coulomb one. It seems that the harmonic potential models better the real situation of experimentally fabricated wires.

\subsection{Density response function}

In this section we use the dynamical density response theory and the fluctuation-dissipation theorem to obtain the static properties. The density response function $\chi(q,\omega)$ is given by \cite{Giuliani05,Renu12}
\be
\label{RPAresponse}
\chi(q,\omega)=
\frac{\chi_{0}(q,\omega)+\lambda\chi_{1}(q,\omega)}{1-\lambda V(q)[\chi_{0}(q,\omega)+\lambda\chi_{1}(q,\omega)]}
\ee
where $\chi_{1}(q,\omega)=\chi_{1}^{se}(q,\omega)+\chi_{1}^{ex}(q,\omega)$ is the first-order correction to the polarizability which includes exchange and self energy contributions. We indicate the order of potential by a $\lambda$ factor. The expression for the non-interacting polarizability is
\be
\chi_{0}(q,\omega)=g_s\sum_{k} \frac{n_k-n_{k+q}}{\omega+\Omega_{k,q}}
\ee
whereas the selfenergy and exchange contributions respectively are given by \cite{Renu12}
\be
\chi_{1}^{se}(q,\omega)=g_s\sum_{k,p}\frac{v(k-p)(n_k-n_{k+q})(n_p-n_{p+q})}{(\omega+\Omega_{k,q})^2}
\label{chi_se}
\ee
and
\be
\chi_{1}^{ex}(q,\omega)=-g_s\sum_{k,p}\frac{v(k-p)(n_k-n_{k+q})(n_p-n_{p+q})}{(\omega+\Omega_{k,q})(\omega+\Omega_{p,q})}.
\label{chi_ex}
\ee
Here $\Omega_{k,q}=\omega_k-\omega_{k+q}$, $\Omega_{p,q}=\omega_p-\omega_{p+q}$, the spin degeneracy factor is $g_s$ and $n_k$ represents the Fermi-Dirac distribution function.

The first order high-density expansion of Eq.(\ref{RPAresponse}) can be written as,
\be
\label{resHDE}
 \chi(q,i\omega)&=&\chi_{0}(q,i \omega)+\lambda\; v(q)\chi_{0}^2(q,i\omega)\nonumber\\
 & &+\lambda\; \chi_{1}^{se}(q,i\omega)+\lambda \;\chi_{1}^{ex}(q,i\omega)
\ee
and will be used in the further calculations.

\subsection{Ground-state energy}
\label{Groundstateenergy}

With the help of the density-density response function and the fluctuation-dissipation theorem, the ground state energy can be obtained in the form \cite{Vinod18a}
\be
\label{gsE}
E_g&=&E_0+\frac{n}{2}\sum_{q\neq 0}V(q)\nonumber\\
& \times& \bigg( -\frac{1}{n\pi}\int^1_0 d\lambda \int^{\infty}_0 \chi(q,i \omega;\lambda)\; d\omega-1\bigg).
\ee
Using Eq. (\ref{resHDE}) in (\ref{gsE}) a simplified form can be given as the sum of kinetic energy of the non-interacting gas $E_0$, exchange energy $E_x$ and the  correlation energy $E_c$ as
\be
E_g=E_0+E_x+E_c,
\ee
where the exchange energy is
\ba
E_x&=&\frac{n}{2}\sum_{q\neq 0}V(q)\bigg( -\frac{1}{n\pi}\int^1_0 d\lambda \int^\infty_0 \chi_0(q,i\omega)d\omega-1\bigg).
\label{Ex_Ener}
\end{align}
The residual energy (i.e. correlation energy) is
\be
E_c&=&\frac{n}{2}\sum_{q\neq 0}V(q)\bigg( -\frac{1}{n\pi}\int^1_0 d\lambda \int^\infty_0 \bigg\{ \lambda\; V(q) \chi_0^2(q,i\omega)\nonumber\\
& &+\lambda\; \chi_1^{se}(q,i\omega)+\lambda\; \chi_1^{ex}(q,i\omega)\bigg\}d\omega\bigg),
\label{Corr_Ener}
\ee
with $n=(k_F\; g_s)/\pi$ being the linear electron number density and $k_F$ is the Fermi wave vector. The static structure factor is
\be
\label{ssf}
S(q)=-\frac{1}{\pi\; n} \int_{0}^{\infty} d\omega\; \chi^{''}(q,\omega)
\ee
where $\chi^{''}(q,\omega)$ is the imaginary part of the density
response function. The integral in (\ref{ssf}) can be re-written using the
contour integration method \cite{Giuliani05} as
\be
\label{ssfRPA}
S(q)=-\frac{1}{\pi\; n} \int_{0}^{\infty} d\omega\; \chi(q,i
\omega).
\ee
The Eq.s (\ref{Ex_Ener})  and (\ref{Corr_Ener}) can be expressed in terms of the static structure factor as
\be
\label{exchange_energy_SSF}
E_x&=&\frac{n}{2}\sum_{q\neq 0}V(q) [S_0(q)-1)],\\
E_c&=&\frac{n}{4}\sum_{q\neq 0}V(q) [S^{d}_1(q)+S_1^{se}(q)+S_1^{ex}(q)]
\label{ec}
\ee
where
\be
\label{noninteractingSSFS0q}
S_0(q)&=& -\frac{1}{n\pi}\int^\infty_0  \chi_0(q,i\omega)d\omega,\\
S_1^{d}(q)&=& -\frac{1}{n\pi}\int^\infty_0  V(q) \chi_0^2(q,i\omega)d\omega,\\
S_1^{se}(q)&=& -\frac{1}{n\pi}\int^\infty_0 \chi_1^{se}(q,i\omega)d\omega,\label{Sse_d}\\
S_1^{ex}(q)&=& -\frac{1}{n\pi}\int^\infty_0 \chi_1^{ex}(q,i\omega)d\omega.
\label{S1}
\ee
The above expressions are provided for clarity as well as to be self contained in this paper.

\section{Analytical Expressions}
\label{AnalyticalExpressions}

\subsection{Structure factor}

In this subsection we present the results for infinitely-thin wire for the cylindrically and harmonically regularized Coulomb potential. For completeness and coherent presentation we provide these results explicitly as these are also being used for the calculation of the correlation energy.

The non-interacting structure factor (\ref{noninteractingSSFS0q}) is obtained using
\begin{eqnarray}
 \chi_{0}(q,i\omega)=\frac{g_{s} m}{2 \pi q} \ln
\bigg[\frac{\omega^2+(\frac{q^2}{2 m}-\frac{q
k_{F}}{m})^2}{\omega^2+(\frac{q^2}{2 m}+ \frac{q k_{F}}{m})^2}\bigg]
\label{NIRF}
\end{eqnarray}
and reads
\be
S_0(x)&=&
    \begin{cases}
    x,& ~ x< 1 \\

    1,              & ~ x> 1
\end{cases}.
\label{S0}
\ee
where we will use $x=q/2k_F$ in the following.
The first-order static structure factor can be written as
\be
S_1(x)=S_1^{d}(x)+S_1^{se}(x)+S_1^{ex}(x).
\label{S1}
\ee
The analytical evaluation of $S_1^{d}(x)$ and $S_1^{ex}(x)$ for an infinitely-thin cylindrical wire has been reported earlier \cite{KM18}. For the harmonic wire we present the results in this paper. The contribution of the selfenergy to the structure factor  $S_{1}^{se}(q,\omega)$ given in Eq. (\ref{Sse_d}) turns out to be zero due to the $\omega$ integration. The details of the calculations are given in appendix A. In the following we present the total result for both wire models in the small $b$ limit.

For the cylindrically regularized Coulomb potential the sum of both corrections $S_1^{d}(x)$ and $S_1^{ex}(x)$ is given by
\ba
\label{SCyl1}
&S^{Cy.}_1(x)\nonumber\\
&=\frac{g_s^2 r_s}{\pi ^2 x}\left \{
\begin{array}{ll}
\zeta(x) \approx 4 x \ln x+ o(x)&, x<1
\cr
\zeta(x)-2 x \ln x \ln e^2 x\approx {1\over 2 x^3}+o(x^{-4})&, x>1
\end{array}
\right .
\end{align}
with
\be
\zeta(x)&=&(x+1)\ln (x+1)\ln \left ( {x^2 e^2\over x+1} \right )
\nonumber\\
&&+(x-1)\ln |x-1|\ln\left ({x^2 e^2\over |x-1|}\right )
\label{z}
\ee
For the harmonically regularized Coulomb potential, we find the result (\ref{SHrl1}) and (\ref{SHrg1}) which after reformulation shows that the first-order correction to structure factor $S_1(x)$ is exactly the same as for cylindrically regularized potentials though the details of calculations
are quite different as seen in the appendix A.
It is interesting to note that both structure factors are independent of the thickness of the wire $b$.
To the best of our knowledge the equivalence of cylindrical and harmonic wire in the $b\rightarrow 0 $ limit is not known before and it is our new finding.

\begin{figure}
\includegraphics[width=8cm]{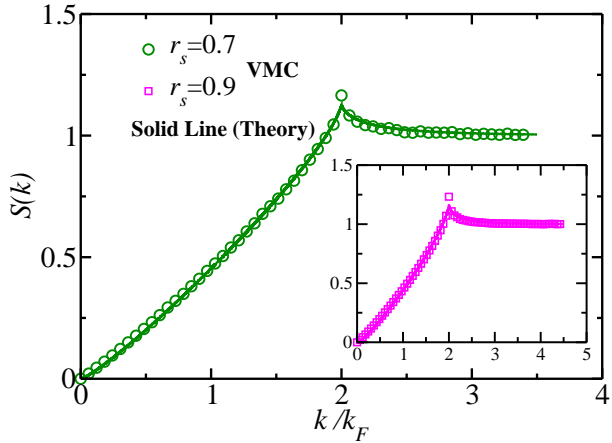}
\caption{\label{ssf_theory_vmc}(Color online) VMC static structure factor (SSF) of an infinitely thin wire with N = 99, compared
with the high-density theory (solid line). The main plot
shows the SSF for rs = 0.7, and the inset is for rs = 0.9.}
\end{figure}

In figure \ref{ssf_theory_vmc} the theoretical structure factor is
plotted alongwith our simulation data \cite{Vinod18c} for
$r_{s}=0.9$ and $0.7$ which are in very good agreement. This shows that the high-density expansion works well up to $r_s\sim 1$.

\subsection{Pair correlation function}
The pair correlation function $g(r)$ is obtained from the static structure factor $S(q)$ as
\begin{eqnarray}
\label{pcftheory}
g(r)=1-\frac{1}{2\pi n}\int^{\infty}_{-\infty} dq\; e^{iqr}[1-S(q)].
\end{eqnarray}

\begin{figure}
\includegraphics[width=8cm]{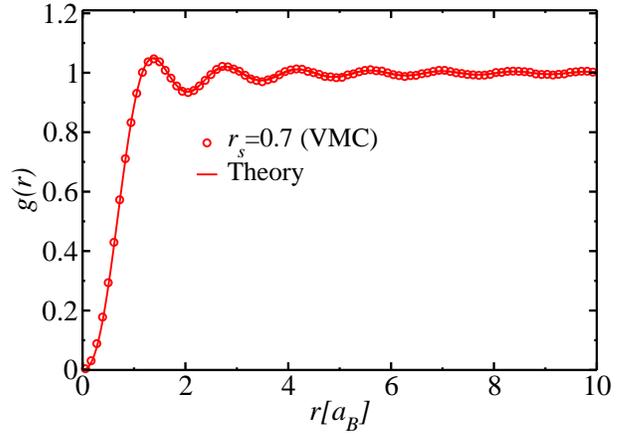}
\caption{\label{gr_p7}(Color online) The pair correlation function  $g(r)$ of 1D homogeneous electron gas (\ref{pcftheory}) in infinitely thin wires. The variational Monte Carlo (VMC) simulation data \cite{Vinod18c} are compared with the high-density expansion (\ref{S1}) at $r_s=0.7$. }
\end{figure}

In figure \ref{gr_p7}, we compare the variational Monte Carlo
simulation with our recent high-density theory \cite{Vinod18c}
formula (\ref{pcftheory}) for infinitely thin wire, which was
obtained in the $b \rightarrow 0$ limit for cylindrical wire. It
is observed that the theory gives a good agreement for small
distances as well as for the oscillations at larger distances.

In the small $r$ limit for infinitely thin wire, the analytical
expression for $g(r)$ can be given for the $r_s\rightarrow 0$
limit  as
\begin{eqnarray}
g(r)=1-\frac{1}{g_s}\bigg( I_1 -\frac{r^2}{2}I_2\bigg),
\end{eqnarray}
where the first two moments are introduced as $I_1= \int^{\infty}_0 (1-S_0(x)-S_1(x)) dx$ and  $I_2= \int^{\infty}_0 x^2 (1-S_0(x)-S_1(x)) dx$. Using $S_0(x)$ and $S_1(x)$ one obtains the analytical values for $I_1$ and $I_2$
and
\begin{eqnarray}
  g(r)= 1-\frac{1}{g_s}+ \frac{{\bar r}^2}{3 g_s} \left(1-\frac{2 \left(3+\pi ^2\right) g_s^2
   r_s }{3\pi^2}\right)+o({\bar r}^4)
\label{new}
\end{eqnarray}
with $\bar r =\frac{r k_F}{\hbar}=\frac{\pi  n r}{g_s}=\frac{\pi}{2 g_S
   r_S}{r\over a_B}$.
This is a new exact result for $g(r)$ at small $r$ in the
$r_s\rightarrow 0$ limit. It is noted that for the ground state of
infinitely-thin and completely polarized ($g_s=1$) wires, the pair
correlation function increases as $r^2$ with reduced positive
curvature compared to the non-interacting case. Further we observe
that the pair correlation function remains only positive for $r_s<
\frac{3}{2} \left(\frac{\pi^2}{\pi^2+3} \right)=1.15$ which sets
the boundary on the applicability of small $r_s$ expansion. Our
results implies that the two-particle wave function approaches
zero linearly at the coalescence point. This is  consistent with
the Kimball identity \cite{Kimball73} and also with the quantum
Monte Carlo simulation wave function.

\begin{figure}
\label{smallpcf_rsp1p2p3}
\includegraphics[width=8cm]{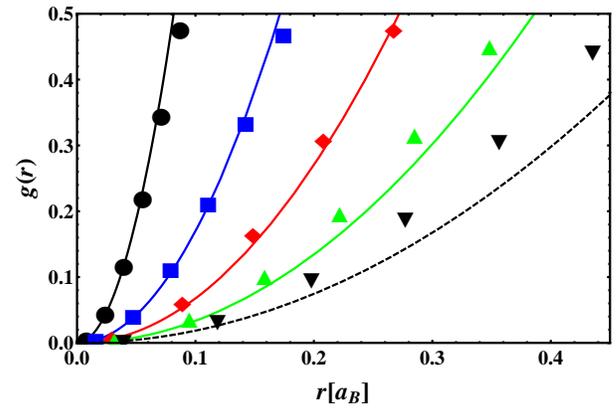}
\caption{\label{gr_small}(Color online) The small $r$ pair correlation function $g(r)$ for $r_s=0.1-0.5$ from left to right where VMC data (points) are compared to formula (\ref{new}) (thick lines).}
\end{figure}

Eq. (\ref{new}) is
plotted in figure \ref{gr_small} together with the VMC simulation
data \cite{Vinod18c}. As one can see the curvature is well reproduced up to $r_s<0.4$ and $r<0.4 a_B$ above which it starts to deviate at larger $r$.

\subsection{Exchange energy}

Substituted Eq. (\ref{S0}) in (\ref{exchange_energy_SSF}) provides the analytical expression of exchange energy for the finite thickness of a harmonic wire. After some simplification one obtains
\ba
\epsilon_x(r_s,p)=&-\frac{(1+p)^2}{16r_s} \bigg\{G_{2,3}^{2,2}\left(\frac{b^2\pi^2}{4r_s^2}(1+p)^2|
\begin{array}{c}
 0,\frac{1}{2} \\
 0,0,-\frac{1}{2} \\
\end{array}
\right)\nonumber\\
 -&\frac{4r_s^2}{b^2\pi^2(1+p)^2}\bigg[\ln\left(\frac{b^2\pi^2}{4r_s^2}(1+p)^2\right)\nonumber\\
 +&\text{exp}\left(\frac{b^2\pi^2}{4r_s^2}(1+p)^2\right)
   \Gamma \left(0,\frac{b^2\pi^2}{4r_s^2}(1+p)^2\right)+\gamma \bigg]\bigg\}\nonumber\\
    -&\frac{(1-p)^2}{16r_s} \bigg\{G_{2,3}^{2,2}\left(\frac{b^2\pi^2}{4r_s^2}(1-p)^2|
\begin{array}{c}
 0,\frac{1}{2} \\
 0,0,-\frac{1}{2} \\
\end{array}
\right)\nonumber\\
 -&\frac{4r_s^2}{b^2\pi^2(1-p)^2}\bigg[\ln\left(\frac{b^2\pi^2}{4r_s^2}(1-p)^2\right)\nonumber\\
+&\text{exp}\left(\frac{b^2\pi^2}{4r_s^2}(1-p)^2\right)
   \Gamma \left(0,\frac{b^2\pi^2}{4r_s^2}(1-p)^2\right)+\gamma \bigg]\bigg\}\nonumber\\
\label{Ex_har}
\end{align}
where $b=\bar b 2k_F$, $G_{2,3}$ is the Meijer G function and $\Gamma$  is incomplete gamma function \cite{Bateman53}, respectively. We use here the polarization $p$ if one integrates up to the Fermi momenta $k_{\uparrow\downarrow}=k_F(1\pm p)/2$.

Similarly the analytical expression for the exchange energy of the cylindrical wire is given by
\ba
 \epsilon_x(r_s,p)=&-\frac{(1+p)^2}{8r_s}\bigg(\frac{\frac{\pi}{2r_s}(1+p)b\;K_1[\frac{\pi}{2r_s}(1+p)b]-1}{2(\frac{\pi}{4r_s}(1+p)b)^2}\nonumber\\
  +&\pi K_0[\frac{\pi}{2r_s}(1+p)b] L_{-1}[\frac{\pi}{2r_s}(1+p)b]\nonumber\\
  +&\pi K_1[\frac{\pi}{2r_s}(1+p)b]\;L_{0}[\frac{\pi}{2r_s}(1+p)b]\bigg)\nonumber\\
 -&\frac{(1-p)^2}{8r_s}\bigg(\frac{\frac{\pi}{2r_s}(1-p)b\;K_1[\frac{\pi}{2r_s}(1-p)b]-1}{2(\frac{\pi}{4r_s}(1-p)b)^2}\nonumber\\
 +&\pi K_0[\frac{\pi}{2r_s}(1-p)b] L_{-1}[\frac{\pi}{2r_s}(1-p)b]\nonumber\\
  +&\pi K_1[\frac{\pi}{2r_s}(1-p)b]\;L_{0}[\frac{\pi}{2r_s}(1-p)b]\bigg)
 \label{Ex_Cyl}
\end{align}
where $K_n(x)$  is n$^{th}$ order modified Bessel function of
second kind, and $L_n(x)$ is modified Struve function \cite{Abramowitz72}.

Expressions for the exchange energy of the one dimensional electron gas for any thickness $b$  of the wire and for a given polarization $p$ are new results in the present investigation. These expressions  reduce to infinitely thin cylindrical and harmonic wires as reported in Eq. (5) of reference \cite{Vinod18c}. The formulae are useful for numerical results of the exchange energy in one-dimensional systems.

\subsection{Correlation energy}

The correlation energy per particle in Eq. (\ref{ec}) in the small-$b$ limit for a cylindrical wire is obtained here as
\be
\label{Int_Corr}
\epsilon_c^{Cy.}
&=&\frac{1}{4r_s}\bigg\{\Lambda_{(x<1)}+\Lambda_{(x>1)} \bigg\},
\ee

where we use the small $b$ expansion of $v(x)$ according to (\ref{v}). The result for $x<1$ is
\ba
\label{lamda_clyl1}
&\Lambda_{(x<1)}=\int^1_0 v(x) [S^{Cy.}_1(x)]_{x<1}\;dx
\nonumber\\
&=\frac{r_s g_s^2}{12\pi^2} \bigg\{42 \zeta (3) \ln \left(\frac{bk_F}{8}\right)
+48 (\ln
   (2)-2) \ln (2) \ln (bk_F)
\nonumber\\
&+48 \left(-2
   \text{Li}_4\left(\frac{1}{2}\right)+\ln ^2(2)+\gamma  \left(\ln
   ^2(2)-\ln (4)\right)+\ln (4)\right)\nonumber\\
   & +42 (\gamma -1) \zeta (3)+\pi
   ^4-4 \log ^3(2) (12+\ln (2))+4 \pi ^2 \ln ^2(2)\bigg\},
\end{align}
and for $x>1$ it is
\ba
\label{lamda_clyg1}
&\Lambda_{(x>1)}=\int^\infty_1 v(x) [S_1^{Cy.}(x)]_{x>1}\; dx
\nonumber\\
&=-\frac{2r_s g_s^2}{\pi^2} \bigg\{\frac{7}{4} \zeta (3) \left(\ln
   \left(\frac{bk_F}{8}\right)+\gamma -1\right)-4\text{Li}_4\left(\frac{1}{2}\right)
\nonumber\\
& +\frac{17 \pi
   ^4}{360}+(\ln (2)-2) \ln (4) \ln
   (bk_F)-\frac{\ln ^4(2)}{6}-2 \ln ^3(2)\nonumber\\
   & +\frac{1}{6} \pi ^2 \ln
   ^2(2)+2 \gamma  \ln ^2(2)+2 \ln ^2(2)+\ln (16)-4 \gamma  \ln
   (2)\bigg\},
\end{align}
where $\zeta(s)$ is the Riemann zeta function and $\text{Li}_n(z)$ is the polylogarithm function \cite{Abramowitz72}. Adding Eq.(\ref{lamda_clyl1}) and (\ref{lamda_clyg1}), major cancellations occur and one obtains the correlation energy as
\be
\label{exact}
\epsilon_c(r_s)=-\frac{\pi^2}{360}.
\ee
We believe that this derivation is new and the obtained correlation energy is in excellent agreement with our variational quantum Monte Carlo simulation\cite{Vinod18c} and conventional perturbation theory\cite{Loos13,Loos16}.

The calculation of the correlation energy for cylindrical wire has been described so far. Exactly the same procedure is followed for the harmonic wire. The details of  calculations are given in Appendix B. By adding Eq. (\ref{lamdaHrl1}) and (\ref{lamdaHrg1}) again major cancellations appear and the final result for the correlation energy in a harmonically regularized potential is the same as Eq. (\ref{exact}). The correlation energy turns out to be identical for cylindrical and for harmonic wires indicating that the way the Coulomb potential is regularized is immaterial at least for these two cases.

\section{Conclusions}
\label{Conclusion}
In the present paper we have obtained the analytical expression for the exchange energy of a harmonic and of a cylindrical wire and a given polarization with any finite thickness. We have also presented the structure factor in the high-density limit for infinitely-thin wires. This provides an analytical expression for the pair correlation function at small distances $r$ consistent with Kimball's identity in the one-dimensional case. The result $g(r)$ indicates some limitation of the $r_s$ expansion. It is also concluded that the correlation energies are identically the same for both wires and agree with the variational quantum Monte Carlo simulation and conventional perturbation theory. It is gratifying to see that the structure factor and the correlation energy are independent of the choice of the electron confining in one-dimensions.

\begin{acknowledgments}
KNP acknowledge the financial support of National Academy of Sciences of India for the award of platinum jubilee fellowship and Humboldt foundation for financial support for Dresden visit. We also thank DFG  for financial assistance to enable us to be together which resulted in finalization of this work. The hospitality of the Institute for Materials Science and Max Bergmann Center of Biomaterials at Dresden University of Technology is kindly acknowledged.
\end{acknowledgments}

\appendix
\section{}

For an infinite-thin wire the contribution of the direct term $S_1^{d}(x)$ can be obtained in the small-$b$ limit for $x<1$  as \cite{Vinod18a,KM18}
\begin{eqnarray}
S_1^{d}(x)&=&-\frac{g_s^2 r_s} {\pi ^2 x} \bigg[ 2 \bigg\{(1-x) \ln (1-x)+(x+1) \ln (x+1)\bigg\} \nonumber\\
&\times &\bigg(-\ln \bigg(\frac{b x}{2}\bigg)-\gamma \bigg)\bigg]
\end{eqnarray}
with $x=q/(2k_F)$ and for $x>1$
\be
S_1^{d}(x)&=&-\frac{g_s^2 r_s}{\pi ^2 x}\bigg[2 \bigg\{(x-1) \ln (x-1)-2 x \ln (x)\nonumber\\
&+ &(x+1) \ln (x+1)\bigg\} \left(-\ln \left(\frac{b x}{2}\right)-\gamma \right)\bigg].
\ee
The detailed calculation of the structure factor for the selfenergy $S_1^{se}(x)$ and the exchange contribution $S_1^{ex}(x)$ is explicitly given in Appendix C.  The contribution of the selfenergy to the structure factor  $S_{1}^{se}(q,\omega)$ given in Eq.(\ref{Sse_d}) turns out to be zero due to the $\omega$ integration. The structure factor for exchange contribution for $x<1$ is given by
\be
S_1^{ex}(x)&=&\frac{g_s^2 r_s}{\pi ^2 x} \bigg[ (x-1) \ln (1-x) \bigg\{2 (\ln (b)+\gamma -1)\nonumber\\
&+&\ln
   \left(\frac{1-x}{4}\right)\bigg\}-(x+1) \ln (x+1)\nonumber\\
&\times &\bigg\{2 (\ln (b)+\gamma -1)+\ln
   \left(\frac{x+1}{4}\right)\bigg\}\bigg]
   \label{Sex_less}
\ee
and for $x>1$
\be
S_1^{ex}(x)&=&\frac{g_s^2 r_s}{\pi ^2 x}\bigg[(1-x) \ln \left(\frac{x-1}{x}\right) \bigg\{2 (\ln
   (b)+\gamma )\nonumber\\
   &+&\ln \left(\frac{1}{4} (x-1)
   x\right)\bigg\}+(x+1) \ln \left(\frac{x}{x+1}\right)\nonumber\\
   &\times&\bigg\{2 (\ln (b)+\gamma )+\ln \left(\frac{1}{4} x
   (x+1)\right)\bigg\}\nonumber\\
   &+&2 \bigg\{(x-1) \ln (x-1)-2 x \ln
   (x)\nonumber\\
   &+&(x+1) \ln (x+1)\bigg\}\bigg].
     \label{Sex_large}
\ee
The contribution of the direct term $S_1^{d}(x)$ can be obtained in the small-$b$ limit for $x<1$  as \cite{Vinod18a}
\ba
S_1^{d}(x)=&-\frac{g_s^2 r_s} {\pi ^2 x} \bigg[ 2 \bigg\{(1-x) \ln (1-x)+(x+1) \ln (x+1)\bigg\} \nonumber\\
\times &(-2\ln (b x)-\gamma )\bigg]
\end{align}
and for $x>1$
\ba
S_1^{d}(x)=&-\frac{g_s^2 r_s}{\pi ^2 x}\bigg[2 \bigg\{(x-1) \ln (x-1)-2 x \ln (x)\nonumber\\
+ &(x+1) \ln (x+1)\bigg\} \left(-2\ln (b x)-\gamma \right)\bigg].
\end{align}
The exchange correction for $x<1$ is given by
\begin{align}
S_1^{ex}(x)=&\frac{g_s^2 r_s}{\pi ^2 x}\bigg[ (x-1) [2 \ln (b)+\gamma -2] \ln (1-x)\nonumber\\
    -&(x+1) \ln
   (x+1) [2 \ln (b)+\ln (x+1)+\gamma -2]\nonumber\\
    +&(x-1) \ln
   ^2(1-x) \bigg]
\end{align}
and for $x>1$
\ba
S_1^{ex}(x)=&\frac{g_s^2 r_s}{\pi ^2 x}\bigg[-(x-1) \ln \left(\frac{x-1}{x}\right) [2 \ln
   (b) +\ln (x-1)\nonumber\\
   +&\ln (x)+\gamma ]+(x+1) \ln
   \left(\frac{x}{x+1}\right) [2 \ln (b)\nonumber\\
   +&\ln (x)+\ln (x+1)+\gamma ]+2 x \ln \left(x^2-1\right)\nonumber\\
   -&4 x \ln (x)+4
   \coth ^{-1}(x) \bigg].
\end{align}

Together the first-order correction to the structure factor $S_1(x)$ is given for a harmonically regularized potential and $x<1$ as
\ba
\label{SHrl1}
S^{Hr.}_1(x)=&\frac{g_s^2 r_s}{\pi ^2 x} \bigg[-(x+1) \ln ^2(x+1)+2
   (x+1)\nonumber\\
 \times & (\ln (x)+1) \ln (x+1) +(x-1) \ln (1-x) \nonumber\\
\times &(\ln
   (1-x)-2 (\ln (x)+1))\bigg]
\end{align}
and for $x>1$ by
\ba
\label{SHrg1}
S^{Hr.}_1(x)=&-\frac{g_s^2 r_s}{\pi ^2 x} \bigg[-2 x (\ln (x)+1) \ln
   \left(x^2-1\right)\nonumber\\
+&(x-1) \ln ^2(x-1)+\ln ^2(x+1)\nonumber\\
+& x
   \left(\ln ^2(x+1)+2 \ln (x) (\ln (x)+2)\right)\nonumber\\
-&4 (\ln
   (x)+1) \coth ^{-1}(x)\bigg].
\end{align}

\section{}
The correlation energy per particle in the small-$b$ limit for harmonic wires is given by
\be
\label{Int_Har}
\epsilon_c^{Hr.}&=&\frac{1}{4r_s}\bigg\{\Lambda_{(x<1)}+\Lambda_{(x>1)} \bigg\}
\ee
for $x<1$
\ba
\label{lamdaHrl1}
&\Lambda_{(x<1)}=\int^1_0 v(x) [S^{Hr.}_1(x)]_{x<1}\;dx
\nonumber\\
&=\frac{r_s g_s^2}{\pi^2} \bigg( \frac{7}{4} \zeta (3) (2 \ln (b)+\gamma -2-4 \ln (2))+4
   \ln ^2(2) \ln (b)\nonumber\\
   &  -8 \ln (2) \ln (b)-8
   \text{Li}_4\left(\frac{1}{2}\right)+\frac{\pi
   ^4}{12}-\frac{\ln ^4(2)}{3}+\frac{1}{3} \pi ^2 \ln
   ^2(2)
\nonumber\\&
+2 \gamma  \ln ^2(2)-\gamma  \ln (16)-4 (\ln
   (2)-2) \ln (2)\bigg)
\end{align}
and for $x>1$
\ba
\label{lamdaHrg1}
&\Lambda_{(x>1)}=\int^\infty_1 v(x) [S^{Hr.}_1(x)]_{x>1}\; dx
\nonumber\\
&=-\frac{r_s g_s^2}{\pi^2} \bigg(   -\frac{7}{4} \zeta (3) (2 \ln (b)+\gamma -2-4 \ln (2))-4
   \ln ^2(2) \ln (b)\nonumber\\
   & +8 \ln (2) \ln (b)+8
   \text{Li}_4\left(\frac{1}{2}\right)-\frac{17 \pi
   ^4}{180}+\frac{\ln ^4(2)}{3}-\frac{1}{3} \pi ^2 \ln
   ^2(2)
\nonumber\\
&-2 \gamma  \ln ^2(2)+\gamma  \ln (16)+4 (\ln
   (2)-2) \ln (2)\bigg).
\end{align}

\section{}
The selfenergy in Eq. (\ref{chi_se}) and the exchange contribution in Eq.(\ref{chi_ex}) can be simplified further as
\ba
\chi_{1}^{se}(q,\iota\omega)&=2g_s\sum_{k,p}n_k n_p [v(k-p)-v(k-p+q)]\nonumber\\
&\times \frac{\Omega^2_{k,q}-\omega^2}{(\Omega^2_{k,q}+\omega^2)^2}
\label{chi_sen}
\end{align}
and
\ba
\chi_{1}^{ex}(q,\iota\omega)&=-2g_s\sum_{k,p}n_kn_p\bigg\{ \frac{v(k-p)(\Omega_{k,q}\Omega_{p,q}-\omega^2)}{(\Omega^2_{k,q}+\omega^2)(\Omega^2_{p,q}+\omega^2)}\nonumber\\
&+\frac{v(k+p+q)(\Omega_{k,q}\Omega_{p,q}+\omega^2)}{(\Omega^2_{k,q}+\omega^2)(\Omega^2_{p,q}+\omega^2)}\bigg\}.
\label{chi_exe}
\end{align}

It may be noted that now $\omega$ is real in the expressions of $\chi_{1}^{se}(q,\omega)$ and $\chi_{1}^{ex}(q,\omega)$. The contribution of the selfenergy to the structure factor  $S_{1}^{se}(q,\omega)$ given in Eq.(\ref{Sse_d}) turns out to be zero due to the $\omega$ integration.

The exchange contribution $\chi_{1}^{ex}(q,\omega)$ in Eq.(\ref{chi_exe}) is further simplified by using the transformation $-p^\prime=p+q$ and then $k=k-q/2$, $p=p-q/2$ leading to
\begin{eqnarray}
\chi_{1}^{ex}(q,\iota\omega)&=&-2g_s\sum_{k,p}\bigg\{v(k\!-\!p) [n_{k-\frac{q}{2}}n_{p-\frac{q}{2}}\!-\!n_{k-\frac{q}{2}}n_{p+\frac{q}{2}}]\nonumber\\
& &\times\frac{(\Omega_{k-\frac{q}{2},q}\;\Omega_{p-\frac{q}{2},q}-\omega^2)}{(\Omega^2_{k-\frac{q}{2},q}+\omega^2)(\Omega^2_{p-\frac{q}{2},q}+\omega^2)}\bigg\}.
\label{chi_exe2}
\end{eqnarray}

Summation over $k,p,\sigma$ is simply written as a sum over $k,p$. This sum over $k,p$ includes both positive and negative values. Using the transformation $k-p=k^{\prime}$ and replacing the sum by integration one can obtain the structure factor as
\ba
&S_{1}^{ex}(q,\iota\omega)=\frac{g_s^2}{n2\pi^3}\int \limits_{0}^{\infty}\int \limits_{0}^{\infty}\int \limits_{0}^{\infty} \mathrm{d} k\,\mathrm{d}p\,\mathrm{d}\omega\; v(k)
\nonumber\\&
\times\bigg\{\left (n_{k\!+\!p\!+\!\frac{q}{2}}n_{p\!+\!\frac{q}{2}}\!-\!n_{k\!+\!p\!+\!\frac{q}{2}}n_{p\!-\!\frac{q}{2}}\right)
\frac{[\frac{p(k\!+\!p)q^2}{m^2}\!-\!\omega^2]}{[\frac{q^2(k\!+\!p)^2}{m^2}\!+\!\omega^2][\frac{q^2p^2}{m^2}\!+\!\omega^2]}
\nonumber\\&
\!+\!\left (n_{k\!-\!p\!+\!\frac{q}{2}}n_{p\!-\!\frac{q}{2}}\!-\!n_{k\!-\!p\!+\!\frac{q}{2}}n_{p\!+\!\frac{q}{2}}\right)\frac{[\!-\!\frac{q^2}{m^2}(k\!-\!p)p\!-\!\omega^2]}{[\frac{q^2}{m^2}(k\!-\!p)^2\!+\!\omega^2][\frac{q^2}{m^2}p^2\!+\!\omega^2]} \nonumber\\
&\!+\!\left (n_{k\!-\!p\!-\!\frac{q}{2}}n_{p\!+\!\frac{q}{2}}\!-\!n_{k\!-\!p\!-\!\frac{q}{2}}n_{p\!-\!\frac{q}{2}}\right)\frac{[\frac{q^2}{m^2}(\!-\!k\!+\!p)p\!-\!\omega^2]}{[\frac{q^2}{m^2}(k\!-\!p)^2\!+\!\omega^2][\frac{q^2}{m^2}p^2\!+\!\omega^2]}\nonumber\\
&\!+\!\left (n_{k\!+\!p\!-\!\frac{q}{2}}n_{p\!-\!\frac{q}{2}}\!-\!n_{k\!+\!p\!-\!\frac{q}{2}}n_{p\!+\!\frac{q}{2}}\right)\frac{[(\!-\!\frac{q}{m})^2(\!-\!k\!-\!p)p\!-\!\omega^2]}{[\frac{q^2(k\!+\!p)^2}{m^2}\!+\!\omega^2][\frac{q^2p^2}{m^2}\!+\!\omega^2]}\bigg\}.\nonumber\\
\label{SSF_ex1}
\end{align}
The $\omega$-integration in Eq. (\ref{SSF_ex1}) can be performed analytically
as well,
\begin{eqnarray}
&&S_{1}^{ex}(q,\iota\omega)=\frac{g_s^2}{n2\pi^3}\int \limits_{0}^{\infty}\int \limits_{0}^{\infty} \mathrm{d} k\,\mathrm{d}p\; v(k)\nonumber\\
&&\times\bigg\{-\frac{m\pi}{kq}[n_{k-p+\frac{q}{2}}n_{p-\frac{q}{2}}-n_{k-p+\frac{q}{2}}n_{p+\frac{q}{2}}]\nonumber\\
&&-\frac{m\pi}{kq}[n_{k-p-\frac{q}{2}}n_{p+\frac{q}{2}}-n_{k-p-\frac{q}{2}}n_{p-\frac{q}{2}}]\bigg\}.
\label{SSF_ex2}
\end{eqnarray}
After rearranging the distribution function it takes the form
\begin{eqnarray}
&&S_{1}^{ex}(q)=\frac{g_s^2 m}{n 2\pi^2} \int\limits_0^{\infty}\int\limits_0^{\infty}dkdp \frac{v(k)}{kq}\nonumber\\
&&\times  [n_{k-p-\frac{q}{2}}-n_{k-p+\frac{q}{2}}][n_{p-\frac{q}{2}}-n_{p+\frac{q}{2}}].
\end{eqnarray}

We use now $[n_{p-\frac{q}{2}}-n_{p+\frac{q}{2}}]=\Theta(p^+-p)\Theta(p-|p^-|)$ and $[n_{k-p-\frac{q}{2}}-n_{k-p+\frac{q}{2}}]=\Theta((p+p^+)-k)\Theta(k-(p+|p^-|))$, where $p^{\pm}=\frac{q\pm 2k_F}{2}$. To prove the identity $[n_{p-\frac{q}{2}}-n_{p+\frac{q}{2}}]=\Theta(p^+-p)\Theta(p-|p^-|)$, we write the distribution function in terms of Heaviside step functions $\left[\Theta(k_F^2-(p-\frac{q}{2})^2)-\Theta(k_F^2-(p+\frac{q}{2})^2) \right]$ with the inequalities $k_F^2-(p-\frac{q}{2})^2>0$ and $k_F^2-(p+\frac{q}{2})^2<0$. The inequality $k_F^2-(p-\frac{q}{2})^2>0$ eventually leads to   $\left( p- \frac{(q+2k_F)}{2}\right)\left( p- \frac{(q-2k_F)}{2}\right)<0$ with the definite range of $p$ i.e. $p<\frac{q+2k_F}{2}$ and $p>\frac{q-2k_F}{2}$. Similarly the inequality $k_F^2-(p+\frac{q}{2})^2<0$ leads to $\left( p- \frac{(-q+2k_F)}{2}\right)\left( p+ \frac{(q+2k_F)}{2}\right)>0$ with $p>\frac{-q+2k_F}{2}$. Hence these inequalities can be written in terms of $\Theta$ function as $\Theta(p^+-p)\Theta(p-|p^-|)$. Similarly the above distribution $[n_{k-p-\frac{q}{2}}-n_{k-p+\frac{q}{2}}]$ can be proven by the same procedure. The above integration can be written after simplification for $x>1$ as,
\begin{eqnarray}
S_{1}^{ex}(q,\omega)=-\frac{g_s^2 r_s}{2\pi^2 x}\int\limits^{|x+1|}_{|x-1|}dt \int\limits^{\frac{t+|x-1|}{2}}_{\frac{t+|x+1|}{2}} \frac{v(\bar{x})}{\bar{x}}d\bar{x}
\end{eqnarray}
and for $x<1$
\begin{eqnarray}
S_{1}^{ex}(q,\omega)=-\frac{g_s^2 r_s}{2\pi^2 x}\int\limits^{1+x}_{1-x}dt \int\limits^{\frac{t+(1-x)}{2}}_{\frac{t+(1+x)}{2}} \frac{v(\bar{x})}{\bar{x}}d\bar{x}.
\end{eqnarray}

\begin{figure}[!t]
\centering
\includegraphics[scale=0.35]{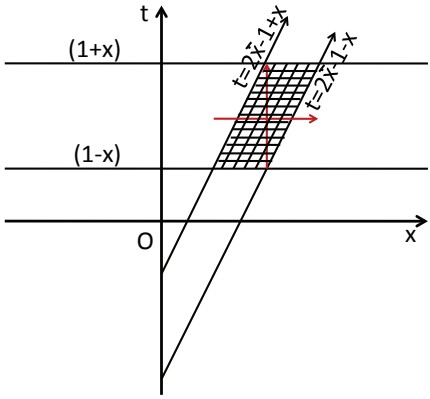}
\caption{(Color online) Graphically interchange the order of integrations from $dt\,d\bar{x}$ to $d\bar{x}\,dt$ for $x<1$. }\label{integral}
\end{figure}

Now we interchange the order of integrations from $dt\,d\bar{x}$ to $d\bar{x}\,dt$ for $x<1$ shown in figure \ref{integral}. We divide the integration in the shaded region in figure \ref{integral} into two parts. The simplified integration after interchanging the order of integration can be written as,
\begin{eqnarray}
S_{1}^{ex}(q,\omega)&=&-\frac{g_s^2 r_s}{2\pi^2 x}\bigg[\int\limits^{1}_{1-x}d\bar{x} \int\limits^{2\bar{x}-1+x}_{1-x} dt \nonumber\\
&+&\int\limits^{1+x}_{1}d\bar{x} \int\limits^{1+x}_{2\bar{x}-1-x} dt \bigg]\frac{v(\bar{x})}{\bar{x}}.
\end{eqnarray}
After performing the $dt$ integration one gets
\begin{eqnarray}
S_{1}^{ex}(q,\omega)&=&-\frac{g_s^2 r_s}{2\pi^2 x}\bigg[\int\limits^{1}_{1-x} (2\bar{x}-2+2 x)d\bar{x}  \nonumber\\
&+&\int\limits^{1+x}_{1}(2+2x-2\bar{x})d\bar{x}  \bigg]\frac{v(\bar{x})}{\bar{x}}.
\end{eqnarray}
The above integration can be written in the form of
\ba
S^{ex}_1(q)=- {g_s^2r_s\over \pi^2 x}
&\left [
\left (
(1+x)\int\limits_{1}^{1+x}-
(1-x)\int\limits_{1-x}^{1}
\right )
{d \bar x\over \bar x} {v}(\bar x)
\right .
\nonumber\\
&\left .
+
\left (
\int\limits_{1-x}^{1}-
\int\limits_{1}^{1+x}
\right )
d {\bar x} v(\bar x)
\right ]
\label{Sv}
\end{align}
and similarly for $x>1$ it yields
\ba
S^{ex}_1(q)=-r_s {g_s^2\over \pi^2 x}
&\left [
\left (
(1+x)\int\limits_{x}^{1+x}-
(x-1)\int\limits_{x-1}^{x}
\right )
{d \bar x\over \bar x} {v}(\bar x)
\right .
\nonumber\\
&\left .
+
\left (
\int\limits_{x-1}^{x}-
\int\limits_{x}^{1+x}
\right )
d {\bar x} v(\bar x)
\right ].
\label{Sva}
\end{align}
The explicit integrals appearing in (\ref{Sv}) and (\ref{Sva}) can be solved analytically and are given in Eqs. (\ref{Sex_less}) and (\ref{Sex_large}).

\end{document}